\documentclass[aps,prl,twocolumn,showpacs,superscriptaddress,groupedaddress]{revtex4}  
\usepackage{graphicx}  
\usepackage{dcolumn}   
\usepackage{bm}        
\usepackage{amssymb}   
\usepackage{amsmath}
\usepackage{xcolor}
\usepackage{booktabs, multirow, array, makecell, caption}

\begin{document}

\widetext

\title{Prediction of the excitation energies of the 2$^+_1$ states for superheavy nuclei based on the microscopically derived Grodzins relation}

\author{N.Yu. Shirikova}
\address{Joint Institute for Nuclear Research, 141980 Dubna, Moscow region, Russia}
\author{A.V. Sushkov}
\address{Joint Institute for Nuclear Research, 141980 Dubna, Moscow region, Russia}
\author{L.A. Malov}
\address{Joint Institute for Nuclear Research, 141980 Dubna, Moscow region, Russia}
\author{E.A. Kolganova}
\address{Joint Institute for Nuclear Research, 141980 Dubna, Moscow region, Russia}
\address{Dubna State University, 141982 Dubna, Moscow Region, Russia}
\author{R.V. Jolos}
\address{Joint Institute for Nuclear Research, 141980 Dubna, Moscow region, Russia}
\address{Dubna State University, 141982 Dubna, Moscow Region, Russia}

\date{\today}

\begin{abstract}
\begin{description}
\item[Background:] As the result of synthesis of nuclei with large proton numbers a new region of investigations of the structure of nuclei has been discovered. Due to the recent significant increase in the yield of superheavy nuclei their gamma-spectroscopic studies become possible.
\item[Purpose:] To predict the excitation energies of the $2^+_1$ states of nuclei with Z$\ge 100$.
\item[Method:] The microscopic variant of the Grodzins relation derived based on the geometrical collective model and a microscopic approach to description of the low-energy nuclear structure is applied.
\item[Results:]  The excitation energies of the $2^+_1$ states of the even-even nuclei from $^{256}$Fm to $^{296}_{120}$X which differ from each other in the number of $\alpha$-particles are predicted.
\item[Conclusion:] It is shown that at the beginning of the chain of the studied nuclei the excitation energies of the $2^+_1$ states don't exceed 100 keV. Then $E(2^+_1)$ sharply increases with $A$ and reaches maximum value of $400-500$ keV in $^{284}$Fl or $^{292}$Og.
\end{description}
\end{abstract}

\maketitle

\section{Introduction}

The synthesis of nuclei with large proton numbers up to Z = 118~\cite{Oganessian1,Oganessian2,Oganessian3,Hofmann,Hofmann1,Morita,Munzenberg}  has lead to discover of a new region for investigations of the structure of nuclei, namely, the investigation of the structure of the superheavy nuclei. A number of interesting experimental results have already been obtained~\cite{Herzberg,Hessberger,Streicher,Samark}
and calculations that provide information on the single particle spectra and evolution of the shape of these nuclei  with increasing Z have been performed
 \cite{Dullmann,Rudolf,Cwiok1,Cwiok2,Bender10,Berger,Parkhomenko1,Parkhomenko2,Sobiczewski10,Moller,Sobiczewski11,Heenen10,Bezbakh10,Shen,Zhou1,Meng,Zhou2,Ma,Long,Lalazissis,Kruppa,Shi,Zhou3,Li,Bender,Warda,BJP,DFT,Gogny,RMF}. A number of calculations of the excited states spectra of superheavy nuclei were made in \cite{Cwiok10,Delaroche10,Heenen20,Egido,Jolos2,Jolos3,Shirikova1,Shirikova2,Kartavenko}. Currently, the main source of experimental information on excitation spectra  of superheavy nuclei is their $\alpha$-decay. With the start of work  of the Factory of superheavy elements in Dubna and due to a significant increase in the yield of such nuclei in fusion reactions, gamma spectroscopic studies  became possible in this area of the nuclide chart.

One of the most interesting questions related to the study of the properties of superheavy nuclei is the question of the next magic number of protons after Z = 82.  It is well known how accurate indicator of what numbers of protons and neutrons are magic, is the behavior of the excitation energy of the $2^+_1$
 states of even-even nuclei. When the values of Z and N approach the magic numbers, the value of $E(2^+_1)$ increases sharply and reaches a maximum at double magic nuclei. This makes information about the excitation energies of the $2^+_1$ states of the even-even superheavy nuclei important for understanding  of their structure. It should be noted that the excitation energy of the $2^+_1$ state also gives, in principle, information  about  the shape of these nuclei.

 For the experiments planned  to measure $E(2^+_1)$ it could be useful to know theoretical predictions of the values of $E(2^+_1)$ in the region of the nuclide chart
 under investigation. The well known  Grodzins relation formulated in 1962 \cite{Grodzins}, which established that the product of the energy of the $2^+_1$ state per probability of the E2 transition from the ground state of the nucleus to the $2^+_1$ state, is a smooth function of A and Z. This property does not depend on whether the nucleus is spherical or deformed, although both $E(2^+_1)$ and
$B(E2;0^+_1\rightarrow 2^+_1)$  vary through a large factor. This is especially important when analyzing the properties of nuclei from those parts of the nuclide chart where the transition from deformed to spherical nuclei occurs.
Later on Raman et al. \cite{Raman1,Raman2} by analyzing a larger set of the experimental data have shown that the Grodzins relation can be presented in the following form
\begin{eqnarray}
\label{Eq1}
E(2^+_1)\times B(E2;0^+_1\rightarrow 2^+_1)=2.57(45)Z^2A^{-2/3}
\end{eqnarray}
where $E(2^+_1)$ is given in keV and $B(E2;0^+_1\rightarrow 2^+_1)$ in $e^2b^2$. This relation has been often used to estimate the unknown $B(E2;0^+_1\rightarrow 2^+_1)$ values from the known $E(2^+_1)$  in different nuclei, especially in nuclei close to the nuclear drip line, since measuring of $B(E2)$ is a much
more demanding task than measuring excitation energies. In the case of superheavy nuclei both quantities presented in the Grodzins relation are unknown.
However, the value of B(E2;$0^+_1\rightarrow 2^+_1$) is directly related to quadrupole deformation of nuclei $\beta_2$. But this quantity was the object of the numerous theoretical calculations, whose results are quite close to each other. This circumstance give us a possibility to use the results of calculations of $\beta_2$ in order to determine the corresponding values of $B(E2;0^+_1\rightarrow 2^+_1)$, and then using the Grodzins relation to predict $E(2^+_1)$.

The value of the proportionality coefficient in the relation (\ref{Eq1}) has been considered in details in \cite{Pritychenko} using the latest set of the experimental data. The analysis indicates on a strong reason for individual fit of the proportionality coefficient in (\ref{Eq1}) for separate groups of nuclei.

In our previous paper \cite{Jolos1} we have derived Grodzins relation based on the collective quadrupole Bohr Hamiltonian and reproduced the $A$-dependence of the Grodzins product. At the same time, in \cite{Jolos1} was proposed a method for calculating the proportionality coefficient in the Grodzins relation, based on the microscopic model of nuclear structure.

The purpose of this work is to calculate the proportionality coefficient in the Grodzins relation for a group of nuclei with $Z\ge 100$, based on a microscopic nuclear model, and to predict on this basis the energies of $2^+_1$ states of even-even superheavy nuclei.

\section{Brief derivation of the Grodzins relation}

Let us repeat shortly a derivation of the Grodzins relation, suggested in \cite{Jolos1}.
It was noted in \cite{Jolos1} that the form of the Grodzins relation indicates that it can be derived using technique of the energy-weighted sum-rule. Therefore, the relation can be derived by analyzing the double commutator of the quadrupole operator $Q_{2\mu}$ with the collective Bohr Hamiltonian
which has the form
\begin{eqnarray}
\label{Eq2}
H=-\frac{\hbar^2}{2}\sum_{\mu,\mu'}\frac{\partial}{\partial\alpha_{2\mu}}\left(B^{-1}\right)_{\mu\mu'}\frac{\partial}{\alpha_{2\mu'}} +V(\alpha_{2\mu}),
\end{eqnarray}
where $\left(B^{-1}\right)_{\mu\mu'}$ is an inverted inertia tensor, $\alpha_{2\mu}$ are the collective variables and  $V$ is the potential. It is convenient to present the inertia tensor in terms of the components having fixed values of the angular momentum $L$,
\begin{eqnarray}
\label{Eq3}
\left(B^{-1}\right)_{\mu\mu'}=\sqrt{5}\sum_{LM}C^{LM}_{2\mu  2\mu'}\left(B^{-1}\right)_{LM},
\end{eqnarray}
where $C^{LM}_{2\mu  2\mu'}$ is a Clebsch Gordan coefficient. For double commutator we obtain
\begin{eqnarray}
\label{Eq4}
[[H,Q_{2\mu}],Q_{2\mu'}]=-\hbar^2 q^2\sqrt{5}\sum_{LM}C^{LM}_{2\mu  2\mu'}\left(B^{-1}\right)_{LM},
\end{eqnarray}
where $q=3/4\pi eZr_0^2 A^{2/3}$.
Taking the average of (\ref{Eq4}) over the ground state $|0^+_1\rangle$ we obtain
\begin{eqnarray}
\label{Eq5}
\sum_n E(2^+_n)\times B(E2;0^+_1\rightarrow 2^+_n)\nonumber\\
=\frac{5}{2}\hbar^2 q^2 \langle 0^+_1|\left(B^{-1}\right)_{00}|0^+_1\rangle,
\end{eqnarray}
where summation takes place over all collective quadrupole $2^+$ states related to the surface mode treated by the Bohr Hamiltonian.
Only one term in (\ref{Eq5}) containing $2^+_1$ state is included in the Grodzins relation. Therefore, it is necessary to evaluate the contribution of the remaining terms in the total sum. In the limit of harmonic quadrupole oscillations E2 transition from the ground state is possible only to the $2^+_1$ state. In this case, leaving on the left in (\ref{Eq5}) only the transition to the $2^+_1$ state, we get on the right  the proportionality factor equal to 5/2. Let us consider the experimental data on spherical nuclei. The probabilities of the E2 transitions from the ground to the first, second and third $2^+$ states are experimentally known in some cases. We transfer the contributions corresponding to the last two transitions to the right part of the equation (\ref{Eq5}).
We get that the proportionality factor on the right side in (\ref{Eq5}) is equal for $^{106}$Pd - 4.7/2, for $^{108}$Pd - 4.8/2 and for $^{112}$Cd - 4.4/2. Thus, the experimental data for spherical nuclei are close to the values for the harmonic limit.
Consider another limiting case, namely, rigid rotational motion. In \cite{Brentano}, it was shown that non-zero contributions to the sum of (\ref{Eq5}) are also given by E2 transitions from the ground state to the $2^+$ states of the beta- and gamma-rotational bands.
Moving the corresponding terms to the right side of the relation (\ref{Eq5}), and leaving on the left only the term corresponding to the transition
$0^+_1\rightarrow 2^+_1$, we get the proportionality coefficient on the right equal to 2/2.
The experimental data for deformed nuclei are as follows. In the case of $^{164}$Dy - 3.9/2, in the case of $^{158}$Gd - 3.6/2. The same consideration of experimental data for nuclei transitional between spherical and deformed gives the following results: $^{154}$Gd - 1.9/2, $^{152}$Sm - 3.5/2, $^{152}$ Gd - 4.4/2, $^{148}$Sm - 2.9/2, $^{196}$Pt - 4.0/2, $^{194}$Pt - 4.9/2, $^{192}$Pt - 4.9/2.
We see that the values of this coefficient show a large variation. In our calculations below, we  use a proportionality factor of 2/2 as it follows theoretically for the limiting case of the deformed nuclei, since according to the numerous calculations, nuclei with Z$=100-110$ considered below are well deformed very heavy nuclei. This is just the value of the proportionality factor that makes it possible to reproduce below the only experimentally known value of $E(2^+_1)$ for one of the nuclei considered below, namely, for $^{256}$Fm.
At the same time our predictions for nuclei with smaller deformation will underestimate the values of  $E(2^+_1)$ for this choice of the proportionality factor.
Thus, below, the following relation will be used to calculate the value of $E(2^+_1)$
\begin{eqnarray}
\label{Eq6}
E(2^+_1)\times B(E2;0^+_1\rightarrow 2^+_1)\nonumber\\
=\frac{2}{2}\hbar^2q^2\langle 0^+_1|\left(B^{-1}\right)_{00}|0^+_1\rangle .
\end{eqnarray}

The components of the inertia tensor given in the laboratory frame can be expressed through their components in the intrinsic frame \cite{Jolos4}:
\begin{eqnarray}
\label{Eq7}
\left(B^{-1}\right)_{LM}&=&D^L_{M0}\left(B^{-1}\right)_{L0}^{in} \nonumber\\
&+&\frac{1}{\sqrt{2}}\left(D^L_{M2}+D^L_{M-2}\right)\left(B^{-1}\right)_{L2}^{in}\nonumber\\
&+&\frac{1}{\sqrt{2}}\left(D^L_{M4}+D^L_{M-4}\right)\left(B^{-1}\right)_{L4}^{in},
\end{eqnarray}
where $\left(B^{-1}\right)_{LK}^{in}$ in general case depends on $\beta$ and $\gamma$. We see that only
$\left(B^{-1}\right)_{00}^{in}$ contributes to (\ref{Eq5}).
In the case of axial symmetry, when we put $\gamma$=0
the components $\left(B^{-1}\right)_{L0}^{in}$ can be expressed through the inertia coefficients for $\beta$- and $\gamma$- motion, and
the rotational inertia coefficient \cite{Jolos4}:
\begin{eqnarray}
\label{Eq8}
\frac{1}{B_{\beta}}=\left(B^{-1}\right)_{00}^{in}-\sqrt{\frac{10}{7}}\left(B^{-1}\right)_{20}^{in}+3\sqrt{\frac{2}{7}}\left(B^{-1}\right)_{40}^{in},\\
\label{Eq9}
\frac{1}{B_{\gamma}}=\left(B^{-1}\right)_{00}^{in}+\sqrt{\frac{10}{7}}\left(B^{-1}\right)_{20}^{in}+\frac{1}{2}\sqrt{\frac{2}{7}}\left(B^{-1}\right)_{40}^{in},\\
\label{Eq10}
\frac{1}{B_{rot}}=\left(B^{-1}\right)_{00}^{in}-\frac{1}{2}\sqrt{\frac{10}{7}}\left(B^{-1}\right)_{20}^{in}-2\sqrt{\frac{2}{7}}\left(B^{-1}\right)_{40}^{in}.
\end{eqnarray}
We obtain from (\ref{Eq8})-(\ref{Eq10}) that
\begin{eqnarray}
\label{Eq11}
\left(B^{-1}\right)_{00}^{in}=\frac{2}{5}\frac{1}{B_{rot}}+\frac{2}{5}\frac{1}{B_{\gamma}}+\frac{1}{5}\frac{1}{B_{\beta}}.
\end{eqnarray}
Substituting (\ref{Eq11}) and the relation $B(E2;0^+_1\rightarrow2^+_1)=q^2\beta^2_2$, which is, in fact, an experimental definition of $\beta_2$, into
(\ref{Eq6}) we obtain
\begin{eqnarray}
\label{Eq13}
E(2^+_1)=\hbar^2\frac{1}{\beta^2_2}\left(\frac{2}{5}\frac{1}{B_{rot}}+\frac{2}{5}\frac{1}{B_{\gamma}}+\frac{1}{5}\frac{1}{B_{\beta}}\right).
\end{eqnarray}

The cranking model expression for the inertia coefficients $B_{\beta}, B_{\gamma}$, and $B_{rot}$ in the case of the single particle Hamiltonian with Woods-Saxon potential are
\begin{widetext}
\begin{eqnarray}
\label{Eq14}
B_{rot}=2\hbar^2R_0^2\sum_{s,t}\frac{|\langle s|\frac{dV}{dr}\frac{1}{\sqrt{2}}(Y_{21}+Y_{2-1})|t\rangle|^2\left(\varepsilon_s\varepsilon_t-(E_s-\lambda)(E_t-\lambda)-\Delta_s\Delta_t\right)}{2\varepsilon_s\varepsilon_t(\varepsilon_s+\varepsilon_t)^3},\\
\label{Eq15}
B_{\gamma}=2\hbar^2R_0^2\sum_{s,t}\frac{|\langle s|\frac{dV}{dr}\frac{1}{\sqrt{2}}(Y_{22}+Y_{2-2})|t\rangle|^2\left(\varepsilon_s\varepsilon_t-(E_s-\lambda)(E_t-\lambda)+\Delta_s\Delta_t\right)(\varepsilon_s+\varepsilon_t)}{2\varepsilon_s\varepsilon_t((\varepsilon_s+\varepsilon_t)^2-\omega_{\gamma}^2)^2},\\
\label{Eq16}
B_{\beta}=2\hbar^2R_0^2\sum_{s,t}\frac{|\langle s|\frac{dV}{dr}Y_{20}|t\rangle|^2\left(\varepsilon_s\varepsilon_t-(E_s-\lambda)(E_t-\lambda)+\Delta_s\Delta_t\right)(\varepsilon_s+\varepsilon_t)}{2\varepsilon_s\varepsilon_t((\varepsilon_s+\varepsilon_t)^2-\omega_{\beta}^2)^2},
\end{eqnarray}
\end{widetext}
where $E_s$ is the single particle energy, $\lambda$ is the Fermi energy, $\varepsilon_s$ is the single quasiparticle energy, $\Delta_s$ is the energy gap parameter depending on the single particle quantum number, and $\omega_{\beta}, \omega_{\gamma}$ are the energies of the $\beta$- and $\gamma$- phonons.
By entering definitions
\begin{eqnarray}
\label{Eq17}
B_x\equiv 2\hbar^2R^2_0\Sigma_x
\end{eqnarray}
where $x=rot, \beta, \gamma$ and substituting (\ref{Eq17}) into (\ref{Eq13}) we obtain
\begin{eqnarray}
\label{Eq18}
E(2^+_1)=\frac{1}{2\beta^2_2R^2_0}\left(\frac{2}{5}\frac{1}{\Sigma_{rot}}+\frac{2}{5}\frac{1}{\Sigma_{\gamma}}+\frac{1}{5}\frac{1}{\Sigma_{\beta}}\right),
\end{eqnarray}
where $\Sigma_{\beta}, \Sigma_{\gamma}$, and $\Sigma_{rot}$ are the sums in Eqs. (\ref{Eq14}) - (\ref{Eq16}).

\section{Model and results}

As it is seen from (\ref{Eq14}) - (\ref{Eq16}) in order to calculate the quantities $\Sigma_{rot}$, $\Sigma_{\gamma}$ and $\Sigma_{\beta}$ presented in the
expression for $E(2^+_1)$ we need in the single particle and single quasiparticle energies, matrix elements of the single particle operators and the energies
of $\beta$- and $\gamma$- vibrations. All these quantities can be calculated in the framework of the Quasiparticle Phonon Model (QPM) \cite{Soloviev1,Soloviev2,Ivanova,Soloviev3,Gareev,Antonenko}. Although this method is not selfconsistent it provides a sufficiently powerfull tool for extensive calculations and predictions.

The Hamiltonian of QPM used below has the following structure
\begin{eqnarray}
\label{Eq19}
H=H_{sp}+H_{pair}+H_M.
\end{eqnarray}
The mean field potential $V_{sp}$ in $H_{sp}$ contains the central potential $V_{WS}$ in the Woods-Saxon form for the neutrons and protons, the spin-orbit part $V_{so}$, and the Coulomb field $V_C$ for protons:
\begin{eqnarray}
\label{Eq20}
V_{sp}=V_{WS} + V_{so}+V_C,
\end{eqnarray}
where
\begin{eqnarray}
\label{Eq21}
V_{WS}=-V_0\left(1+\exp[(r-R(\theta,\varphi)/a]\right)^{-1}.
\end{eqnarray}
Here the axially deformed form of the Woods-Saxon potential is assumed
\begin{eqnarray}
\label{Eq22}
R(\theta,\varphi)=R_0\left(1+\beta_0+\sum_{\lambda =2,4}\beta_{\lambda}Y_{\lambda,0}(\theta,\varphi)\right),
\end{eqnarray}
where $R_0=r_0A^{1/3}$. The parameters of the potential are given in \cite{Adamian}.

The term $H_{pair}$ describes the monopole pairing forces with the strength set to reproduce the odd-even mass differences. After Bogoliubov transformation,
 we obtain the Hamiltonian in terms of the quasiparticle creation and annihilation operators.

 The term $H_M$ in (\ref{Eq19}) describes the multipole interaction of quasiparticles. The separable forces expressed through the operators of the
 multipole moments are used as the residual interaction
 \begin{equation}
\label{Eq23}
H_M=-\frac{1}{2}\sum_{l,\mu}\sum_{\tau,\rho_{\tau}=\pm 1}\left(\kappa_0^{(l\mu)}+\rho_{\tau}\kappa_1^{(l\mu)}\right)M^+_{l\mu}(\tau)M_{l\mu}(\tau)
\end{equation}
Here $\tau$ denotes neutrons or protons. The isoscalar $\kappa_0^{(l\mu)}$ and isovector $\kappa_1^{(l\mu)}$ constants depend on angular momentum and projection $\mu$ on the symmetry axis. The choice of their values was justified in \cite{Soloviev2,Malov1,Shirikova2}. We have used the set of interaction constants suggested for the region of heavy nuclei \cite{Malov1}.
$H_M$ generates phonon excitations in nuclei.

Using single particle wave functions and single particle  energies of the Hamiltonian $ H _ {SP} $ with Woods-Saxon potential,  u-v coefficients of the Bogoliubov transformation, the values of $\Delta _ {\nu} $  and the single quasiparticle energies calculated taking into account $ H _ {pair} $, as well as the energies of the
$\beta $ - and $\gamma $ -- vibrations
obtained using $ H _ M $, we can calculate $\Sigma_{rot}$, $\Sigma_{\gamma}$ and $\Sigma_{\beta}$ necessary to find $E(2^+_1)$ using the relation (\ref{Eq18}).

Below, we use the relation (\ref{Eq18}) to predict the energies of $2^+_1$ states of a series of superheavy nuclei with proton numbers from 100 to 120. As an example, we take a chain of even-even nuclei from $ ^ {256} _ {100} $Fm to $ ^ {296} _ {120} $X, which differ from each other in the number of $\alpha $-particles.
According to numerous calculations, these nuclei include both highly deformed nuclei and  spherical (or close to spherical) nuclei. Due to this reason for our purpose it is convenient to use the Grodzins relation, since its use is not limited to any particular shape of nuclei. The practice of using it in the case of well studied nuclei has confirmed it.

Table I shows the results of calculations of the energies of the first $2^+_1$ states of nuclei from the selected chain.

\begin{table*}[tbh]
\centering
\setlength\aboverulesep{0pt}\setlength\belowrulesep{0pt}
\setcellgapes{3pt}\makegapedcells
\caption{The predicted energies of the $2^+_1$ states. Calculations are based on the microscopic variant of the Grodzins relation (\ref{Eq18}). The variants [A] and [B] are calculated with the parameters obtained using Strutinsky procedure and the single particle level scheme described in the text. These variants differ in the values of parameters of the spin-orbit part of the single particle potential. The variants [Kowal] and [M\"oller] are calculated with deformation parameters taken from \cite{Kowal} and \cite{Moller1}, correspondingly.}
\label{table:one}
\begin{tabular}{l||c|c||c|c||c|c||c|c }
\hline
Nucleus  & $\beta_2$ [A] & $E(2^+_1)$ (keV) & $\beta_2$ [B] & $E(2^+_1)$ (keV) & $\beta_2$ [Kowal] & $E(2^+_1)$ (keV) & $\beta_2$ [M\"oller] & $E(2^+_1)$ (keV) \\
\hline
$^{256}$Fm & 0.279 & 44 & 0.266 & 49 &  0.25 & 58 & 0.240 & 23  \\
$^{260}$No & 0.287 & 42 & 0.267 & 49 &  0.25 & 53 & 0.242 & 71  \\
$^{264}$Rf & 0.275 & 43 & 0.249 & 51 &  0.24 & 61 & 0.232 & 57  \\
$^{268}$Sb & 0.263 & 34 & 0.252 & 37 &  0.23 & 39 & 0.232 & 23  \\
$^{272}$Hs & 0.231 & 75 & 0.235 & 72 &  0.23 & 73 & 0.221 & 40  \\
$^{276}$Ds & 0.232 & 89 & 0.224 & 95 &  0.21 & 100& 0.210 & 49  \\
$^{280}$Cn & 0.181 & 86 & 0.180 & 87 &  0.19 & 83 & 0.086 & 247 \\
$^{284}$Fl & 0.139 & 217& 0.173 & 141&  0.15 & 140& 0.064 & 473 \\
$^{288}$Lv &-0.137 & 202&-0.137 & 185&-0.12 & 256& 0.075 & 453 \\
$^{292}$Og & 0.083 & 532& 0.074 & 523& 0.08 & 433& 0.075 & 196 \\
$^{296}$120&-0.102 & 176&-0.102 & 168& 0.09 & 324& 0.075 & 161 \\

\hline
\end{tabular}
\end{table*}

\begin{figure}[hbt]
\includegraphics[width=0.45\textwidth]{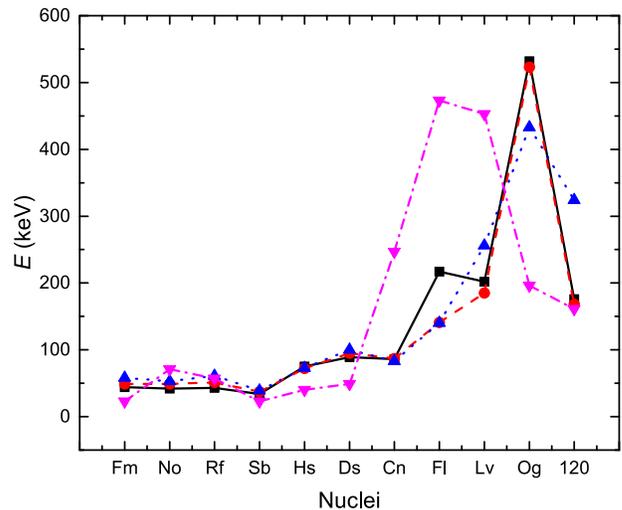}
\caption{\label{f001}The predicted energies of the $2^+_1$ states for different nuclei. Calculations are based on the microscopic variant of the Grodzins relation (\ref{Eq18}) with different sets of quadrupole deformation. Solid line with squares (black) - variant [A], dashed line with circles (red) - variant [B], dot line with triangles (blue) - variant [Kowal], dash-dot line with inverted triangles (magenta) - variant [M\"oller].}
\end{figure}

\begin{figure}[hbt]
\includegraphics[width=0.45\textwidth]{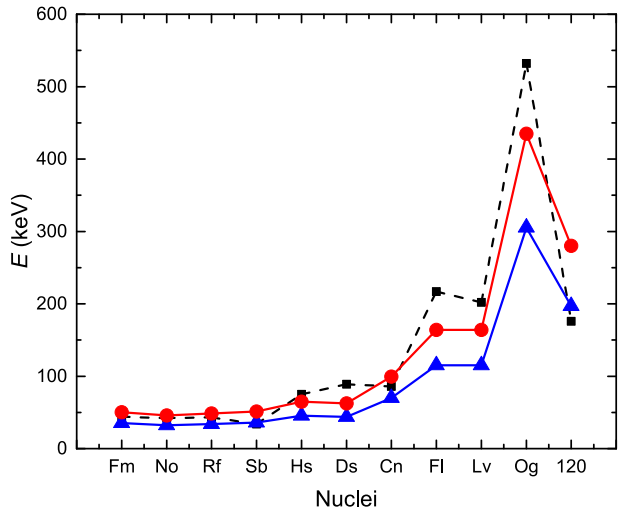}
\caption{\label{f002} The predicted energies of the $2^+_1$ states for different nuclei. Calculations are performed for the microscopic variant of the Grodzins relation (\ref{Eq18}) (variant [A]) and the phenomenological Grodzins relation (\ref{Eq1}) ($E(2^+_1)_{max}$ and $E(2^+_1)_{min}$). Solid line with circles (red) - $E(2^+_1)_{max}$, solid line with triangles (blue) - $E(2^+_1)_{min}$, dashed line with squares (black) - variant [A].}
\end{figure}

Columns $2-5$ show the results of  calculations based on the Hamiltonian QPM with Woods-Saxon potential described above and quadrupole deformations obtained using the Strutinsky method
\cite{Strutinsky1,Strutinsky2}. Thus, these deformations correspond to a minimum of the potential energy surface. The two variants of the calculations indicated in Table I as [A] and [B] differ in the values of the parameters of the spin-orbital part of the Woods-Saxon potential. The results presented in columns 6-9 are obtained using the same single particle level scheme as above, however, with the values of the quadrupole deformation $\beta_2$ taken from \cite{Kowal} (denoted as [Kowal]) and from \cite{Moller1} (denoted as [M\"oller]). Thus, in these two cases equilibrium deformations are not determined selfconsistently with the single particle level scheme used for calculations of the inertia coefficient. However, this give us a possibility to check a sensitivity of the obtained results to variations in deformation parameters.

The results of calculations are illustrated also in Fig.1. We see that qualitatively all four variants given in Table I are similar in the nature of the change with Z. At the beginning of the studied region at Z = $100-110$, where quadrupole deformation is large, the energies of $2 ^ + _ 1 $ states do not exceed 100 keV, i.e. correspond to rotational states. Then, with  decrease in deformation, $E(2 ^ + _ 1)$ rises sharply and reaches a maximum value of $400-500$ keV  in $ ^ {284} $Fl ([M\" oller]) or in $ ^ {292} $Og ([A], [B], [Kowal]), i.e. in nuclei with minimal value of $\beta_2$. Clearly, this reflects the characteristics of the underlying single particle level scheme.
As noted above, our predictions for nuclei with small deformation underestimate the values of $E(2^+_1)$. Thus, values of $400-500$ keV should be considered as the lower boundary.
Note that even in nuclei with Z = $114-120$ the number of neutrons is far from the magic one $N=184$.
Note also that the use of the above microscopic model to calculate the characteristics of nuclei with a very small deformation requires justification. This work, however, does not consider this issue.

In Fig. 2 are  shown the results of the calculations obtained using equation (\ref{Eq18}), as well as the results obtained using the phenomenological formula (\ref{Eq1}). Since in (\ref{Eq1}) the coefficient on the right side of the equation  also contains information about the error of its definition, in Fig. 2 are shown two lines corresponding to the maximum $E(2^+_1)_{max}$ and minimum
$E(2^+_1)_{min}$ values of this coefficient. In the calculations  a set of deformation parameters $\beta_2$ given in the second column of Table I is used. It is seen from Fig. 2  that the results of calculations based on  (\ref{Eq18})  go beyond the limits bounded by the  lines corresponding to $E(2^+_1)_{max}$ and $E(2^+_1)_{min}$ in more than half of cases.

\section{Conclusion}

Based on the  Grodzins relation derived using the Bohr collective Hamiltonian and the microscopical model of nuclear structure,  the excitation energy of the first $2^+$ states of the chain of even-even superheavy nuclei with Z from 100 to 120 are predicted. Calculations are performed for several sets of deformation parameter $\beta_2$ calculated by us or taken from the other publications. We see that for all sets of deformation parameters,  at the beginning of the studied region of nuclei at Z = $100-110$, where quadrupole deformation is large, the energies of the $2^+_1$ states do not exceed 100 keV, i.e. correspond to rotational states. Then, with  decrease in  deformation, $E(2^+_1)$ rises sharply and reaches a maximum value of $400-500$ keV  in $^{284}$Fl  or in
$^{292}$Og, i.e. in nuclei with minimal value of $\beta_2$.

\section{Acknowledgments}
The authors  acknowledge support by the Ministry
of Education and Science (Russia) under Grant No. 075-10-2020-117.



\begin{thebibliography}{99}

\bibitem{Oganessian1} Yu. Ts. Oganessian, J. Phys. G {\bf 34}, R165 (2007).
\bibitem{Oganessian2} Yu. Ts. Oganessian, F. S. Abdullin, P. D. Bailey, D. E. Benker,
M. E. Bennett, S. N. Dmitriev, J. G. Ezold, J. H. Hamilton,
R. A. Henderson, M. G. Itkis {\em et al.}, Phys. Rev. Lett. {\bf 104}, 142502 (2010).
\bibitem{Oganessian3} Yu. Ts. Oganessian, F. S. Abdullin, S. N. Dmitriev, J. M. Gostic,
J. H. Hamilton, R. A. Henderson, M. G. Itkis, K. J. Moody,
A. N. Polyakov, A. V. Ramayya {\em et al.}, Phys. Rev. C {\bf 87}, 014302 (2013).
\bibitem{Hofmann} S. Hofmann, D. Ackermann, S. Antalic, H. G. Burkhard, V. F.
Comas, R. Dressler, Z. Gan, S. Heinz, J. A. Heredia, F. P.
Hessberger {\em et al.}, Eur. Phys. J. A {\bf 32}, 251 (2007).
\bibitem{Hofmann1} S. Hofmann, Rev. Mod. Phys. {\bf 72}, 733 (2000).
\bibitem{Morita} K. Morita, K. Morimoto, D. Kaji, H. Haba, K. Ozeki et al., J. Phys. Soc. Jap. {\bf 81}, 103201 (2012).
\bibitem{Munzenberg} G. M\"unzenberg, Nucl.Phys. A {\bf 944}, 5 (2015).
\bibitem{Herzberg} R.-D. Herzberg and P. T. Greenlees, Prog. Part. Nucl. Phys. {\bf 61},
674 (2008).
\bibitem{Hessberger} F. P. Hessberger, Eur. Phys. J. D {\bf 45}, 33 (2007).
\bibitem{Streicher} B. Streicher, F.P. Hessberger, S. Antalic, S. Hofmann, D. Ackermann, S. Heinz, B. Kindler, J. Khuyagbaatar, I. Kojouharov, P. Kuusiniemi, M. Leino, B. Lommel, R. Mann, S. Saro, B. Sulignano, J. Uusitalo, M. Venhart, Eur. Phys. J. A {\bf 45}, 275 (2010).
\bibitem{Samark} A. Samark-Roth, D. M. Cox, D. Rudolf, I. G. Sarmiento, B. G. Carlsson et al., Phys. Rev. Lett. {\bf 126}, 032503 (2021).
\bibitem{Dullmann} {\it Special Issue on Superheavy Elements}, edited by
Ch. E. D\"ullmann, R.-D. Herzberg, W. Nazarewicz, and Y.
Oganessian, Nucl. Phys. A {\bf 944}, 1 (2015).
\bibitem{Rudolf} D. Rudolph (Ed.), EPJ Web Conf. {\bf 131} (2016).
\bibitem{Cwiok1} S. Ćwiok, S. Hofmann, and W. Nazarewicz, Nucl. Phys. A {\bf 573},
356 (1994).
\bibitem{Cwiok2} S. Ćwiok, W. Nazarewicz, and P. H. Heenen, Phys. Rev. Lett.
{\bf 83}, 1108 (1999).
\bibitem{Bender10} M.Bender, K. Rutz, P. G. Reinhardt, J. A. Maruhn, W. Greiner, Phys. Rev. C {\bf 60}, 034304 (1999).
\bibitem{Berger} J. E. Berger, I. Bitaud, J. Decharge, M.Girod, K. Dietrich, Nucl. Phys. A {\bf 685}, 1 (2001).
\bibitem{Parkhomenko1} A. Parkhomenko and A. Sobiczewski, Acta Phys. Pol. B {\bf 35},
2447 (2004).
\bibitem{Parkhomenko2} A. Parkhomenko and A. Sobiczewski, Acta Phys. Pol. B {\bf 36},
3115 (2005).
\bibitem{Sobiczewski10} A. Sobiczewski, K. Pomorski, Prog. Part. Nucl. Phys. {\bf 58}, 292 (2007).
\bibitem{Moller} P. M\"oller, J. R. Nix, W. D. Myers, and Swiatecki, At. Data Nucl. Data Tables {\bf 59}, 185 (1995).
\bibitem{Sobiczewski11} A.Sobiczewski, Radiochim. Acta {\bf 99}, 395 (2011).
\bibitem{Heenen10} P.-H. Heenen, J. Skalski, A. Staszczak, D. Vretenar, Nucl.Phys. A {\bf 944}, 415 (2015).
\bibitem{Bezbakh10} A. N. Bezbakh, V. G. Kartavenko, G. G. Adamian, N. V. Antonenko, R. V. Jolos, and V. O. Nesterenko, Phys. Rev. C {\bf 92}, 014329 (2015).
\bibitem{Shen} S. Shen, H. Liang, J. Meng, P. Ring, and S. Zhang, Phys. Rev. C
{\bf 96}, 014316 (2017).
\bibitem{Zhou1} S.-G. Zhou, J. Meng, and P. Ring, Phys. Rev. C {\bf 68}, 034323
(2003).
\bibitem{Meng} J. Meng, K. Sugawara-Tanabe, S. Yamaji, and A. Arima, Phys.
Rev. C {\bf 59}, 154 (1999).
\bibitem{Zhou2} S.-G. Zhou J. Meng, and P. Ring, Phys. Rev. Lett. {\bf 91}, 262501
(2003).
\bibitem{Ma} Z.-Y.Ma, J. Rong, B.-Q. Chen, Z.-Y. Zhu, andH.-Q. Song, Phys.
Lett. B {\bf 604}, 170 (2004).
\bibitem{Long} W.-H. Long, N. Van Giai, and J. Meng, Phys. Lett. B {\bf 640}, 150
(2006).
\bibitem{Lalazissis} G. A. Lalazissis, J. Konig, and P. Ring, Phys. Rev. C {\bf 55}, 540
(1997).
\bibitem{Kruppa} A. T. Kruppa, M. Bender, W. Nazarewicz, P.-G. Reinhard, T.
Vertse, and S. Cwiok, Phys. Rev. C {\bf 61}, 034313 (2000).
\bibitem{Shi} Y. Shi, D. E. Ward, B. G. Carlsson, J. Dobaczewski, W.
Nazarewicz, I. Ragnarsson, and D. Rudolph, Phys. Rev. C {\bf 90},
014308 (2014).
\bibitem{Zhou3} S.-G. Zhou, Phys. Scr. {\bf 91}, 063008 (2016).
\bibitem{Li} Z.-X. Li, Z.-H. Zhang, and P.-W. Zhao, Front. Phys. {\bf 10}, 102101
(2015).
\bibitem{Bender} M. Bender, P.-H. Heenen, and P.-G. Reinhard, Rev. Mod. Phys.
{\bf 75}, 121 (2003).
\bibitem{Warda} M. Warda and J. L. Egido, Phys. Rev. C {\bf 86}, 014322 (2012).
\bibitem{BJP} D. Bonatsos, I.E. Assimakis, N. Minkov, A. Martinou, S.K. Peroulis, S. Sarantopoulou, R.B. Cakirli, R.F. Casten, K. Blaum,  Bulg. J. Phys. {\bf 44}, 385 (2017).
\bibitem{DFT} S.E. Agbemava, A. V. Afanasjev, T. Nakatsukasa, and P. Ring, Phys. Rev. C {\bf 92}, 054310 (2015.
\bibitem{Gogny} J.-P. Delaroche, M. Girod, J. Libert, H. Goutte, S. Hilaire, S. Péru, N. Pillet, and G. F. Bertsch, Phys. Rev. C {\bf 81}, 014303 (2010).
\bibitem{RMF} G.A. Lalazissis, S. Raman, and P. Ring, At. Data Nucl. Data Tables {\bf 71} (1999).
\bibitem{Cwiok10} S. Cwiok, P.-H. Heenen, W. Nazarewicz, Nature {\bf 433}, 705 (2005).
\bibitem{Delaroche10} J.-P.Delaroche, M. Girod, H. Goutte, J. Libert, Nucl. Phys. A {\bf 771}, 103 (2006).
\bibitem{Heenen20} P.-H. Heenen, B. Bally, M. Bender, W. Ryssens, Eur. Phys. J Web Conf. {\bf 131}, 02001 (2016).
\bibitem{Egido} J. L. Egido, A. Jungclaus, Phys. Rev. Lett. {\bf 126}, 192501 (2021).
\bibitem{Jolos2} R. V. Jolos, L. A. Malov, N. Yu. Shirikova, and A. V. Sushkov, J. Phys. G {\bf 38}, 115103 (2011).
\bibitem{Jolos3} R. V. Jolos, N. Yu. Shirikova, and A. V. Sushkov,  Phys. Rev. C {\bf 86}, 044320 (2012).

\bibitem{Shirikova1}  N. Yu. Shirikova, A. V. Sushkov, R. V. Jolos,  Phys. Rev. C {\bf 88}, 064319 (2013).
\bibitem{Shirikova2}  N. Yu. Shirikova, A. V. Sushkov, L. A. Malov, and R. V. Jolos,  Eur. Phys. J. A {\bf 51}, 21 (2015).
\bibitem{Kartavenko} V. G. Kartavenko, N. V. Antonenko, A. N. Bezbach, L. A. Malov, N. Yu. Shirikova, A. V. Sushkov, and R. V. Jolos, Chin. Phys. C {\bf 41}, 074105 (2017).

\bibitem{Grodzins}L. Grodzins, Phys. Lett. {\bf 2}, 88 (1962).
\bibitem{Raman1} S. Raman, C.W. Nestor, Jr., K. H. Bhatt, Phys. Rev. C {\bf 37}, 805 (1988).
\bibitem{Raman2} S. Raman, C.W.Nestor, T. Tikkanen, At. Data Nucl. Data Tables, {\bf 78}, 1 (2001).

\bibitem{Pritychenko} B. Pritychenko, M. Birch, B. Singh, Nucl. Phys. A {\bf 962}, 73 (2017).

\bibitem{Jolos1} R. V. Jolos, E. A. Kolganova, Phys. Lett. B {\bf 820}, 136581 (2021).


\bibitem{Brentano} R. V. Jolos, P. von Brentano, N. Pietralla, Phys. Rev. C {\bf 71}, 044305 (2005).

\bibitem{Jolos4} R. V. Jolos, P. von Brentano, Phys. Rev. C {\bf 76}, 024309 (2007).


\bibitem{Soloviev1} V. G. Soloviev, {\it Theory of Complex Nuclei} (Pergamon Press, Oxford, UK, 1976)
\bibitem{Soloviev2} V. G. Soloviev, {\it Theory of Atomic Nuclei: Quasiparticles and Phonons} (Institute of Physics Publishing, Bristol, 1992).
\bibitem{Ivanova} S. P. Ivanova, A. L. Komov, L. A. Malov, V. G. Soloviev, Phys. Part. Nucl. {\bf 7}, 450 (1976).
\bibitem{Soloviev3} V. G. Soloviev, A. V. Sushkov, and N. Yu. Shirikova, Phys. Part. Nucl. {\bf 27}, 667 (1996).
\bibitem{Gareev} F. A. Gareev, S. P. Ivanova, and B. N. Kalinkin, Izv. AN SSSR, Ser.Fiz. {\bf 33}, 1690 (1968).
\bibitem{Antonenko} N. V. Antonenko and L. A. Malov, Bull. Rus. Acad. Sc. Physics {\bf 78}, 1137 (2014).

\bibitem{Adamian} G. G. Adamian,  N. V. Antonenko, L. A. Malov, and R.V.Jolos, Phys. Rev. C {\bf 97}, 034308 (2018).
\bibitem{Malov1} L. A. Malov, and V. G. Soloviev, Phys. Part. Nucl. {\bf 11}, 111 (1980).

\bibitem{Strutinsky1} V. M. Strutinsky, Sov. J. Nucl. Phys. {\bf 3}, 149 (1966).
\bibitem{Strutinsky2} V. M. Strutinsky, Nucl. Phys. A {\bf 95}, 420 (1976).
\bibitem{Kowal} M. Kowal, P. Jachimowicz, J. Skalski, arXiv:1203.5013.
\bibitem{Moller1} P. M\"oller, A. J. Sirk, T. Ichikawa, H. Sagawa, At. Data Nucl. Data Tables {\bf 109-110}, 1 (2016).








\end{thebibliography}
\end{document}